\documentclass[pdflatex,sn-basic]{sn-jnl}  
\usepackage[utf8]{inputenc}
\usepackage{booktabs}
\usepackage{amsmath}

\begin{document}

\title[Show and Tell: Prompt Strategies for Style Control in Multi-Turn LLM Code Generation]{Show and Tell: Prompt Strategies for Style Control in Multi-Turn LLM Code Generation}

\author[1]{\fnm{Jeremiah} \sur{Bohr}}\email{bohrj@uwosh.edu}

\affil[1]{\orgdiv{Department of Information Systems}, \orgname{University of Wisconsin Oshkosh}, \orgaddress{\city{Oshkosh}, \state{Wisconsin}, \country{USA}}}

\abstract{Language models generate functionally correct code that tends to be overly verbose, with elaborate documentation and defensive patterns that exceed task requirements. This “verbosity drift” can increase maintenance costs and obscure core logic, even when code passes tests. We study how prompt design shapes stylistic control in multi-turn code generation, focusing on the trade-off between concision and robustness during both initial generation and subsequent enhancement. Using Gemini 2.5 Pro, we compare four prompting strategies across 160 two-turn code-generation sessions for a Python CLI task: a Control condition with minimal task description, an Examples condition with few-shot demonstrations, an Instructions condition emphasizing minimal code that works, and a Combined condition that pairs instructions with examples. All conditions maintained near-ceiling functional accuracy while producing distinct style profiles. At Turn 1, all three interventions reduced verbosity relative to Control, with the Combined condition yielding the largest compression and Examples offering a balanced reduction. At Turn 2, after a standardized enhancement prompt, example-based generation inflated token counts substantially, whereas directive-based prompts preserved compression discipline. Combined exhibited the strongest overall discipline, pairing large initial gains with moderate expansion. Secondary metrics showed that directive-based prompts fostered higher defensive ratios and lower documentation density, consistent with terse but robust implementations, while example-based prompts preserved documentation and modularity. These findings suggest that style control and correctness are separable properties of prompting mechanisms, that examples excel at shaping initial form, and that concise instructions provide the most stable stylistic control in this two-turn workflow.}


\maketitle

\section{Introduction}\label{sec:intro}
Language models generate functionally correct code that is substantially more verbose than necessary. Excessive documentation, layered control flow, and defensive scaffolding obscure core logic and inflate maintenance effort even when tests pass. Syntheses of LLM use in software engineering show that code generation remains one of the less reliable capabilities, with performance varying markedly across tasks \citep{zheng2025towards,alharbi2026automatic}. As AI coding assistants reshape development work \citep{sarkar2025ai}, the problem is not only whether models produce working code, but whether their stylistic footprint can be constrained.

Code verbosity matters. Inflated code slows review, masks control flow, and increases surface area for faults. Style-focused analyses show that test-passing LLM programs differ systematically from human code in conciseness, documentation density, and robustness practices \citep{wang2025beyond}. Bug taxonomies identify characteristic LLM failure modes, indicating that ``functionally correct'' snapshots may still be brittle \citep{tambon2025bugs}. Surveys report widespread deployment of coding assistants but uneven trust and concern about maintainability \citep{sergeyuk2025using}. Verbosity is not cosmetic; it shapes how developers experience and manage AI-authored code.

Prompts can control code style independent of correctness. In-context demonstrations steer models through formatting patterns even when input--label mappings are corrupted \citep{min2022rethinking}. Context-aware prompting for program repair shows that structured prompts substantially change repair effectiveness without altering the underlying model \citep{li2025context}. Backdoor attacks reveal that carefully chosen exemplars can implant hidden behaviors while preserving test-passing status \citep{qu2025badcodeprompt}. Prompting mechanisms systematically influence style, not only pass rates.

Most evidence comes from single-turn evaluations. In practice, developers use multi-turn workflows to iteratively refine solutions before integration \citep{hao2024empirical}. Enhancement requests add features, harden edge cases, and adapt implementations, creating opportunities for stylistic drift as compact solutions accrete checks and boilerplate. Multi-turn studies of code generation track correctness degradation and error accumulation rather than style changes \citep{li2025beyond,shinn2023reflexion}. Work on long-context reliability documents attention limitations that erode earlier guidance \citep{liu2024lost,laban2025llms,wang2025agents}. Whether stylistic control mechanisms effective on static benchmarks remain effective as code expands is unresolved.

This study tests whether instruction-based, example-based, and combined prompts can impose and maintain compression discipline in multi-turn code generation. We manipulate system prompts across four conditions: Control (no guidance), Instructions (explicit directives), Examples (minimalist demonstrations), and Combined (both). We hold model, decoding parameters, and task specification constant. Functional correctness is controlled via a binary test; verbosity and related stylistic indicators are primary outcomes across 160 paired observations (40 per condition).

We address three research questions: 

\begin{itemize}
    \item {RQ1 (effectiveness):} How do instruction-based, example-based, and combined prompts differ in their initial stylistic control of verbosity relative to a control condition?
    \item {RQ2 (mechanism):} Does combining instructions and examples produce additive, synergistic, or redundant effects on code style?
    \item {RQ3 (expansion discipline):} Which prompts maintain compression discipline when expanding functionality?
\end{itemize}

Our contribution is empirical and conceptual. Empirically, we provide a controlled, multi-turn comparison of prompt strategies for code style under a fixed correctness criterion, quantifying their effects on verbosity, modularity, and defensive practices at both generation and enhancement time. Conceptually, we frame stylistic control as a distinct dimension of reliability for LLM-assisted programming, clarify how different prompting mechanisms influence compression discipline as code grows, and situate these findings within broader evidence on LLM use in software engineering \citep{zheng2025towards}. By focusing on how prompt strategies shape the form of code that developers must read, revise, and maintain across turns, we aim to complement task-level syntheses that emphasize where LLMs succeed or fail and to inform practitioners seeking predictable, controllable style rather than pass rates alone.

\section{Literature Review}\label{sec:lit-review}

\subsection{The Separability of Code Style from Functional Correctness}\label{sec:lit-separability}
Recent scholarship establishes that code generated by large language models (LLMs) exhibits measurable stylistic signatures that differ from human-written code, with differences in size, complexity, readability, conciseness, robustness, naming conventions, and other style attributes even when implementations pass tests or resemble human solutions \citep{fagadau2024analyzing,wang2025beyond,xu2025code_transformed}. Comparative studies use static analysis, manual annotation, and metric suites to quantify non-functional properties beyond pass/fail correctness, including complexity, maintainability, security, robustness, and fine-grained coding style inconsistencies, while contrasting LLM outputs with human implementations or human-oriented baselines \citep{fagadau2024analyzing,wang2025beyond,xu2025code_transformed}. Broader mappings of LLMs in software engineering position code generation as one task within a wider landscape that includes summarization, translation, vulnerability detection, and code evaluation and argue for evaluation beyond simple pass rates, while systematic reviews of automatic code generation emphasize diverse paradigms and validation metrics \citep{zheng2025towards,alharbi2026automatic}.

Beyond taxonomic classification, stylistic differences between human and LLM-generated code are quantifiable: metric-based and style-focused studies apply size, complexity, maintainability, and documentation measures, together with coding-style taxonomies over naming, formatting, control flow, and fault-tolerance, to compare outputs under matched specifications and reveal systematic differences in length, documentation, and robustness-related practices \citep{fagadau2024analyzing,wang2025beyond,xu2025code_transformed}. Recent evaluation of software development agents on SWE-Bench Verified extends this perspective by comparing agent-generated patches to human gold patches using reliability, security, and maintainability indicators, showing that even test-passing patches can diverge structurally, increase complexity, or over-modify the codebase \citep{chen2025evaluating}. Work on design-pattern recognition, bug taxonomies, and structured bug-report summarization likewise shows that LLM-derived embeddings and multi-facet representations can classify structural patterns and recurrent defect modes, including misinterpreted requirements, hallucinated objects or non-existent fields, and missing corner cases, supporting the use of quantitative, multi-dimensional metrics to study stylistic properties as dependent variables \citep{pandey2025design,tambon2025bugs,zhang2026brmds}.

Critically, these non-functional properties of code are separable from functional correctness, typically defined as passing a specified test suite or matching a reference implementation, and several studies show that LLM-generated solutions can satisfy tests while diverging markedly from human baselines in structure, verbosity, and documentation, with such divergences persisting even when language and algorithmic choices are held constant \citep{fagadau2024analyzing,della2025prompt,xu2025code_transformed}. Complementary work on efficiency and evaluation design demonstrates that, even among equally correct solutions, models differ substantially in runtime behavior and in how performance-oriented prompts affect them, while confidence-based and mutation-guided test selection strategies uncover different failure modes and yield different pass rates than naive sampling \citep{niu2024evaluating,quan2025evaluation}. Taken together, this evidence shows that “functionally correct” snapshots can conceal stylistic, structural, and efficiency variation that matters for maintainability and robustness, and justifies treating these non-functional qualities as distinct objects of analysis that can be modeled alongside, but not reduced to, correctness.

\subsection{Mechanisms of Stylistic Control: Instructions versus Exemplars}\label{sec:lit-mechanisms}
Prompt engineering is the standard lever for shaping LLM outputs, with surveys and benchmarks cataloging techniques ranging from zero-shot instructions and few-shot exemplars to chain-of-thought scaffolding, tool calling, multi-turn interaction patterns, and agentic orchestration \citep{qin2024infobench,li2025beyond,wang2025agents}. In software engineering, these techniques are used to specify coding conventions, tailor documentation style, and guide the level of detail in generated code, often without altering the underlying model or decoding strategy. Empirical work on prompt format and stylistic framing shows that declarative instructions and in-context setups can induce systematic shifts in tone, verbosity, and discourse structure in natural language tasks, while studies of code generation report analogous effects on comment density, naming schemes, and defensive or fault-tolerance patterns \citep{min2022rethinking,liu2024lost,della2025prompt,wang2025beyond}. Context-aware prompting for program repair further demonstrates that structured prompts which explicitly stage reasoning, surface relevant context, and constrain the form of patches can significantly improve repair success compared to naive instructions, highlighting prompt design as a mechanism for steering both content and form \citep{li2025context}.

Directives leverage models’ instruction-following capabilities, strengthened by alignment methods such as reinforcement learning from human feedback that tune models to satisfy user-specified behavioral constraints \citep{ouyang2022training}. In code generation, directive-style prompts that request simplified structure, improved readability, or greater robustness measurably shift size, complexity, comment density, and fault-tolerance patterns, with studies reporting redundancy reductions, trade-offs between conciseness and readability or robustness, and modest but consistent changes in correctness and quality metrics across combinations of prompt-programming techniques \citep{wang2025beyond,xu2025code_transformed,khojah2025impact}. Context-aware prompting for program repair further shows that combining high-level instructions with staged reasoning and explicit context selection improves repair success and changes the distribution of generated patches without modifying model parameters, while instruction-following benchmarks document that directive effects are sensitive to wording, position, and competing instructions, indicating that instruction-based control is powerful but uneven across settings \citep{li2025context,liu2024lost,hwang2025llms}.

Exemplars steer model outputs through in-context learning that privileges surface regularities and local structures: few-shot prompts composed of representative input–output pairs can strongly influence formatting conventions, ordering, and local patterns, often with more stable effects than comparable verbal instructions \citep{perez2021true,min2022rethinking,lou2024large}. In software engineering settings, example-based prompts and prompt-pattern designs are used to stabilize structured output formats and documentation patterns, and studies of few-shot example selection in code synthesis show that different exemplars can substantially alter model behavior under fixed tasks \citep{xu2024does,ahlgren2025assisting}. Security-oriented work on poisoned demonstrations further shows that backdoor attacks implemented purely through few-shot examples can reliably implant hidden behavioral patterns in generated code while leaving test-passing rates largely unchanged, underscoring that example-driven control can operate largely orthogonally to correctness \citep{qu2025badcodeprompt}.

Interactive and memory-augmented systems extend single-shot prompting by treating user feedback and prior outputs as additional control channels, as in one-shot correction architectures for natural-language-to-code translation that decompose user requests into sub-problems, construct final programs from user-validated or model-generated snippets, and persist correction data so that edits guide future outputs without retraining \citep{le2024rethinking}. Empirical evaluations of multi-agent and auto-prompting repair pipelines show that LLM agents can self-generate feedback and refine declarative specifications across iterations, often outperforming single-pass baselines \citep{alhanahnah2025empirical}. Related retrieval- and agent-based systems in specialized domains, including UI-to-HTML translation and spacecraft control software, combine domain-specific context with structured prompts to improve code quality and adherence to domain conventions, illustrating that stylistic control can be distributed across prompt design, retrieval or context mechanisms, and feedback loops rather than residing in a single prompt \citep{yuan2025ui2html,he2026enhancing}.

Directives and exemplars therefore constitute distinct pathways for stylistic control: directives specify abstract objectives for the model to operationalize, whereas exemplars supply concrete surface patterns for imitation. Evidence across natural-language and code tasks indicates that directive-based control can influence higher-level organization, robustness, and defensive practices, while exemplar-based control is especially effective at anchoring local form, with interactive and retrieval-augmented workflows extending these influences across iterations \citep{min2022rethinking,le2024rethinking,yuan2025ui2html,li2025beyond,qu2025badcodeprompt,he2026enhancing}. However, existing work has not systematically compared these mechanisms under a fixed correctness criterion or tracked their effects as code grows through enhancement cycles, leaving open questions about redundancy, complementarity, and stability when directives and examples are combined.

\subsection{Compression Discipline During Iterative Enhancement}\label{sec:lit-compression}
Real-world development is iterative rather than single-shot, and analyses of developer–ChatGPT chats embedded in GitHub pull requests and issues show that roughly one-third of linked conversations are multi-turn, with follow-up and refinement prompts together accounting for nearly half of all turns, indicating that developers routinely negotiate solutions across several exchanges rather than a single request–response pair \citep{hao2024empirical}. Within these longer sequences, experimental studies document position effects in which early instructions receive progressively less attention as new content is appended, producing drift in tone, structure, and verbosity across turns \citep{saito2023verbosity,liu2024lost}. Software engineering studies of LLM-assisted editing likewise find that repeated re-prompts and edits can accumulate inconsistencies in naming conventions, documentation style, and defensive patterns, particularly when prompts omit explicit format or retention constraints, making the preservation of stylistic control during enhancement a distinct challenge from establishing it in an initial response \citep{cruz2025prompt,hwang2025llms}.

Evidence from multi-turn interaction research further suggests that maintaining stylistic control during enhancement is challenging because additional turns create opportunities for both necessary and unnecessary expansion: requests that add features, handle new edge cases, or introduce error handling naturally increase code size, yet models also tend to elaborate existing logic, add redundant comments, and introduce auxiliary abstractions that are not required to satisfy updated requirements \citep{saito2023verbosity,xu2024does,laban2025llms,sirdeshmukh2025multichallenge}. Debugging and repair pipelines that repeatedly call LLMs similarly report growing prompts and outputs as more context and scaffolding are accumulated to mitigate earlier failures \citep{alhanahnah2025empirical,li2025beyond}. Across these settings, surveys and empirical studies highlight that stylistic drift depends on how much prior context is preserved, how feedback is phrased, and whether prompts explicitly reassert style constraints \citep{laban2025llms,li2025beyond,sirdeshmukh2025multichallenge}.

Expansion during enhancement is not necessarily evidence of control failure: requests targeting readability, maintainability, or robustness may legitimately increase verbosity through added documentation, modularization, or stronger defensive checks, so the key distinction is between necessary expansion that improves quality and undisciplined drift that inflates code without clear benefit. Multi-dimensional analyses of LLM-generated code and related software artifacts highlight that quality involves distinct facets such as clarity, traceability, and defect coverage, and studies of test selection for LLM-generated programs show that richer suites often reveal latent defects in seemingly minimal solutions, indicating that additional checks and structure are sometimes required for acceptable robustness \citep{quan2025evaluation,tambon2025bugs,zhang2026brmds}. In parallel, work on in-context learning and multi-turn evaluation demonstrates that adding explicit format or length constraints can reduce uncontrolled elaboration, yet such controls are rarely applied directly to style metrics such as token count, modularity, or documentation density \citep{min2022rethinking,laban2025llms,sirdeshmukh2025multichallenge}.

Multi-turn interaction is not solely a source of degradation; surveys of multi-turn generation and frameworks such as self-reflection and execution- or test-feedback loops show that models can use linguistic or runtime signals to correct errors and strengthen outputs across turns \citep{shinn2023reflexion,dakhel2024effective,li2025beyond}. In code generation, dual-agent repair pipelines and retrieval-augmented systems iteratively propose and critique patches or incorporate relevant code and documentation, refining solutions and improving fault revelation without changing model weights \citep{alhanahnah2025empirical,he2026enhancing,yuan2025ui2html}. Survey work on LLM-based agents in software engineering emphasizes perception, memory, and tool use as core components of more autonomous workflows, while one-shot correction architectures in NL-to-code translation demonstrate how persistent feedback and reusable corrections can shift effort from initial implementation to reviewing and shaping model output \citep{wang2025agents,le2024rethinking}. Collectively, these refinement protocols show that models can maintain or improve correctness, robustness, and domain conformance across turns when provided explicit guidance, but they rarely make compression or stylistic discipline a primary target.

These findings converge on an empirical gap: existing work shows that style is separable from correctness and that prompts can steer initial outputs, but it remains unclear which prompting strategies maintain compression discipline when code must expand to satisfy new requirements. This study tests whether instruction-based directives, example-based demonstrations, or their combination differ in both initial effectiveness and expansion behavior. The experimental design manipulates prompting mechanisms across four conditions, measures stylistic outcomes while holding functional correctness constant, and follows initial generation with an enhancement request that simulates realistic feature- and robustness-driven pressure.

\section{Methods}\label{sec:methods}

\subsection{Design and Data Collection}\label{sec:methods-design}
We evaluated four prompting strategies across 160 code-generation sessions using Gemini 2.5 Pro. The design manipulates only the system prompt while holding the task specification, model parameters, and two-turn protocol constant.

Each condition included 10 seeds $\times$ 4 runs per seed $=$ 40 independent observations. Recent work demonstrates that single-prompt evaluations produce brittle results and that evaluation robustness requires aggregation across multiple prompt variants \citep{mizrahi2024state}. Following this principle, seeds diversified sampling and runs within seeds used unique seed values (base\_seed $\times$ 1000 + run\_number) to ensure independent generations. The unit of analysis is the individual run.

\subsection{Prompt Conditions}\label{sec:methods-prompts}
Four between-subjects conditions varied the system prompt:

\textit{Control}. Empty system prompt, establishing baseline code generation without stylistic guidance.

\textit{Instructions}. Explicit directives: ``I value minimal, functional code. No defensive coding unless explicitly required. No docstrings unless function purpose is non-obvious from the name and signature. Write the minimum code that works.''

\textit{Examples}. Two minimalist Python functions demonstrating concise style without explicit rules (full text in Appendix).

\textit{Combined}. Instructions followed by Examples, testing whether directives and demonstrations produce additive effects.

All four conditions received identical user prompts requesting a Python CLI tool for CSV validation and statistical analysis. The task specified required functionality (directory scanning, column validation, statistics calculation, JSON output) and constraints (accept command-line arguments, no test data generation, no cleanup operations). This ensured that observed differences in code style reflected the system prompt manipulation rather than task interpretation.

\subsection{Task and Two-Turn Procedure}\label{sec:methods-task}
All conditions received an identical task prompt requesting a Python CLI tool that scans a directory for CSV files, validates required columns (timestamp, user\_id, event\_type, value), calculates statistics (total events, unique users, average value per event type), and writes results as JSON to an output directory.

Turn 1 generated initial code from this task specification. Turn 2 presented the model with its own Turn 1 output and requested: ``Review and improve this code for readability, maintainability, and error handling.'' The revision instruction was identical across conditions to isolate how initial system prompts affect expansion behavior.

Code was extracted from model responses using regex to identify Python code blocks, selecting the largest block when multiple were present.

\subsection{Measures: Code-Style Outcomes}\label{sec:methods-measures}
We computed four primary metrics from generated code using the \texttt{tiktoken} \texttt{cl100k\_base} encoder and Python's \texttt{ast} module:

\textit{Token count}. Total tokens in the generated program, measuring overall verbosity.

\textit{Defensive ratio}. (try blocks + None checks + assertions) / code lines, measuring error-handling density.

\textit{Documentation density}. Docstrings on functions or classes / (total functions + classes), measuring inline documentation.

\textit{Functions per file}. Total function definitions / number of files, measuring decomposition.

Metrics were computed per run by analyzing all \texttt{.py} files in the output directory. Files with syntax errors received deterministic zero values for AST-derived metrics (functions, classes, docstrings, try blocks, assertions) while retaining token counts and line-based measures.

\subsection{Functional Accuracy}\label{sec:methods-accuracy}
Each program was scored as functionally correct (1) or incorrect (0) using a single binary test. We executed the program with a fixed five-row CSV containing the required columns and accepted success if either: (a) a non-empty JSON file appeared in the output directory, or (b) the program printed summary statistics to \texttt{stdout}. We tried four command-line interface patterns (positional arguments, \texttt{--input}/\texttt{--output} flags, \texttt{--input-dir}/\texttt{--output-dir} flags, \texttt{-i}/\texttt{-o} short flags) to accommodate interface variation. Programs passing any variant received a score of 1. This deliberately minimal functional test isolates stylistic differences under near-ceiling correctness rather than evaluating robust error handling or edge-case coverage.

\subsection{Analytical Strategy}\label{sec:methods-analysis}
\textit{Turn 1}. We compared conditions using one-way ANOVA for each outcome, with pairwise contrasts against Control computed via Welch's t-test. Effect sizes are reported as Cohen's $d$ using pooled standard deviations.

\textit{Turn 2}. Within-condition changes from Turn 1 to Turn 2 were tested using paired t-tests with Cohen's $d$ for paired samples (mean difference / SD of differences). Between-condition differences in change magnitude were tested via one-way ANOVA on delta scores with pairwise contrasts against Control.

All tests are two-sided with $\alpha = .05$. We report unadjusted $p$-values and foreground effect sizes as primary interpretive measures. The sample size ($N = 40$ per condition) provides 80\% power to detect standardized mean differences of $d \approx 0.64$ in pairwise comparisons at $\alpha = .05$.

\section{Results}\label{sec:results}
We analyzed code outputs across four conditions: Control (no system prompt), Instructions (minimal code principles), Examples (concrete code samples), and Combined (principles plus examples). Each condition included 40 independent samples at Turn 1 (initial generation) and Turn 2 (revision after feedback).

``Turn 1: Initial Code Generation''
Functional accuracy was near-perfect across all conditions (317 of 320 samples passed, 99.1\%). With functional requirements met, we examined stylistic outcomes.

Table 1 presents descriptive statistics by condition. Token count varied dramatically by condition. Control produced the most verbose code (M = 2560, SD = 219). Instructions reduced verbosity substantially (M = 1113, SD = 141, d = -7.84, $p < .001$), as did Combined (M = 759, SD = 76, d = -10.97, $p < .001$). Examples showed moderate reduction (M = 2036, SD = 176, d = -2.63, $p < .001$).

\begin{table}[htbp]
\centering
\footnotesize
\caption{Turn 1 Descriptive Statistics by Condition}
\label{tab:descriptives}
\begin{tabular}{lcccc}
\toprule
\textbf{Metric} & \textbf{Control} & \textbf{Instructions} & \textbf{Examples} & \textbf{Combined} \\
& \textbf{M(SD)} & \textbf{M(SD)} & \textbf{M(SD)} & \textbf{M(SD)} \\
\midrule
Token Count & 2560 (219) & 1113 (141) & 2036 (176) & 759 (76) \\
Defensive Ratio & 0.0491 (0.0068) & 0.0640 (0.0134) & 0.0483 (0.0080) & 0.0728 (0.0092) \\
Doc Density & 0.996 (0.023) & 0.379 (0.280) & 0.995 (0.022) & 0.280 (0.268) \\
Functions/File & 3.17 (0.68) & 1.99 (0.46) & 4.41 (0.97) & 2.21 (0.45) \\
\bottomrule
\end{tabular}
\end{table}

Figure 1 displays effect sizes for pairwise contrasts against Control. Defensive coding patterns followed similar trends. Instructions (M = 0.064, SD = 0.013, d = 1.40, $p < .001$) and Combined (M = 0.073, SD = 0.009, d = 2.92, $p < .001$) increased defensive ratio relative to Control (M = 0.049, SD = 0.007). Examples showed no difference from Control (M = 0.048, SD = 0.008, d = -0.11, p = .632).

\begin{figure}[htbp]
\centering
\includegraphics[width=0.95\textwidth]{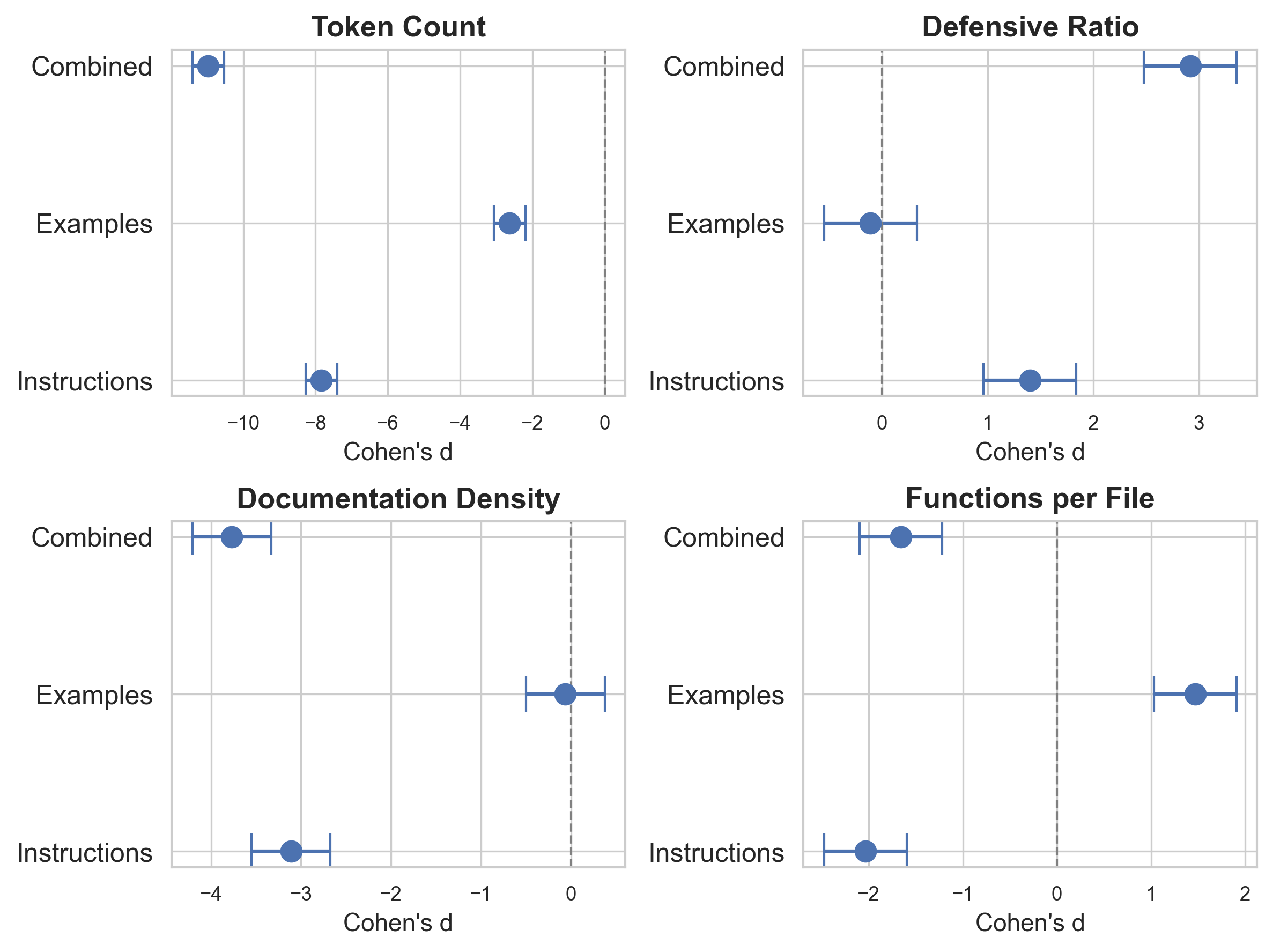}
\caption{Turn 1 Baseline Differences with Control.}
\label{fig:coefficients}
\end{figure}

Documentation density dropped sharply in conditions without examples. Control and Examples maintained near-complete documentation (M = 0.996 and M = 0.995 respectively). Instructions (M = 0.379, SD = 0.280, d = -3.11, $p < .001$) and Combined (M = 0.280, SD = 0.268, d = -3.77, $p < .001$) showed substantial reductions.

Function decomposition patterns diverged by condition. Examples produced the most functions per file (M = 4.41, SD = 0.97, d = 1.47, $p < .001$ vs. Control). Instructions (M = 1.99, SD = 0.46, d = -2.04, $p < .001$) and Combined (M = 2.21, SD = 0.45, d = -1.66, $p < .001$) reduced decomposition relative to Control (M = 3.17, SD = 0.68). Table 2 summarizes all pairwise contrasts.

\begin{table}[htbp]
\centering
\caption{Pairwise Contrasts vs Control (Turn 1)}
\label{tab:pairwise}
\begin{tabular}{lccc}
\toprule
\textbf{Metric} & \textbf{Instructions $d$ ($p$)} & \textbf{Examples $d$ ($p$)} & \textbf{Combined $d$ ($p$)} \\
\midrule
Token Count 
& $-7.84$ ($< 0.001^{***}$) 
& $-2.63$ ($< 0.001^{***}$) 
& $-10.97$ ($< 0.001^{***}$) \\
Defensive Ratio 
& $1.40$ ($< 0.001^{***}$) 
& $-0.11$ ($0.632$) 
& $2.92$ ($< 0.001^{***}$) \\
Doc Density 
& $-3.11$ ($< 0.001^{***}$) 
& $-0.06$ ($0.776$) 
& $-3.77$ ($< 0.001^{***}$) \\
Functions/File 
& $-2.04$ ($< 0.001^{***}$) 
& $1.47$ ($< 0.001^{***}$) 
& $-1.66$ ($< 0.001^{***}$) \\
\bottomrule
\end{tabular}
\end{table}

``Turn 2: Revision Effects''
All conditions increased verbosity when prompted to revise for "readability, maintainability, and error handling." Table 3 presents within-condition changes from Turn 1 to Turn 2.

\begin{table}[htbp]
\centering
\caption{Within-Condition Changes: Turn 1 to Turn 2 (Paired $t$-tests)}
\label{tab:revisions}
\begin{tabular}{llrrrr}
\toprule
\textbf{Outcome} & \textbf{Condition} & $\boldsymbol{\Delta}$ & $\boldsymbol{t}$ & $\boldsymbol{p}$ & $\boldsymbol{d}$ \\
\midrule
Token Count 
& Control      & +266   & 14.12 & $< 0.001^{***}$ & 2.23 \\
& Instructions & +175   & 11.66 & $< 0.001^{***}$ & 1.84 \\
& Examples     & +268   & 16.68 & $< 0.001^{***}$ & 2.64 \\
& Combined     & +126   & 14.54 & $< 0.001^{***}$ & 2.30 \\
\midrule
Defensive Ratio 
& Control      & +0.0025 &  2.58 & $0.014^{*}$    & 0.41 \\
& Instructions & +0.0175 &  5.19 & $< 0.001^{***}$& 0.82 \\
& Examples     & +0.0043 &  3.17 & $0.003^{**}$   & 0.50 \\
& Combined     & +0.0193 &  5.25 & $< 0.001^{***}$& 0.83 \\
\midrule
Documentation Density 
& Control      & +0.020  &  0.78 & 0.440          & 0.12 \\
& Instructions & +0.159  &  2.34 & $0.024^{*}$    & 0.37 \\
& Examples     & -0.008  & -1.43 & 0.160          & -0.23 \\
& Combined     & +0.117  &  1.75 & 0.087          & 0.28 \\
\midrule
Functions/File 
& Control      & +1.90   &  9.00 & $< 0.001^{***}$& 1.42 \\
& Instructions & +1.12   &  7.58 & $< 0.001^{***}$& 1.20 \\
& Examples     & +1.27   &  8.17 & $< 0.001^{***}$& 1.29 \\
& Combined     & +0.97   &  9.35 & $< 0.001^{***}$& 1.48 \\
\bottomrule
\multicolumn{6}{l}{\textit{Note.} $^{*}p<0.05$, $^{**}p<0.01$, $^{***}p<0.001$.}
\end{tabular}
\end{table}

Token count increased significantly in every condition: Control (+266 tokens, t = 14.12, $p < .001$, d = 2.23), Instructions (+175, t = 11.66, $p < .001$, d = 1.84), Examples (+268, t = 16.68, $p < .001$, d = 2.64), and Combined (+126, t = 14.54, $p < .001$, d = 2.30).

Defensive ratio increased across all conditions: Control (+0.0025, t = 2.58, p = .014, d = 0.41), Instructions (+0.0175, t = 5.19, $p < .001$, d = 0.82), Examples (+0.0043, t = 3.17, p = .003, d = 0.50), and Combined (+0.0193, t = 5.25, $p < .001$, d = 0.83).

Documentation density showed inconsistent revision effects. Control showed no change ($\Delta$ = +0.020, t = 0.78, p = .440). Instructions increased ($\Delta$ = +0.159, t = 2.34, p = .024, d = 0.37). Examples showed no change ($\Delta$ = -0.008, t = -1.43, p = .160). Combined showed marginal increase ($\Delta$ = +0.117, t = 1.75, p = .087).

Functions per file increased substantially in all conditions: Control (+1.90, t = 9.00, $p < .001$, d = 1.42), Instructions (+1.12, t = 7.58, $p < .001$, d = 1.20), Examples (+1.27, t = 8.17, $p < .001$, d = 1.29), and Combined (+0.97, t = 9.35, $p < .001$, d = 1.48). Figure 2 illustrates these trajectories across turns.

\begin{figure}[htbp]
\centering
\includegraphics[width=0.95\textwidth]{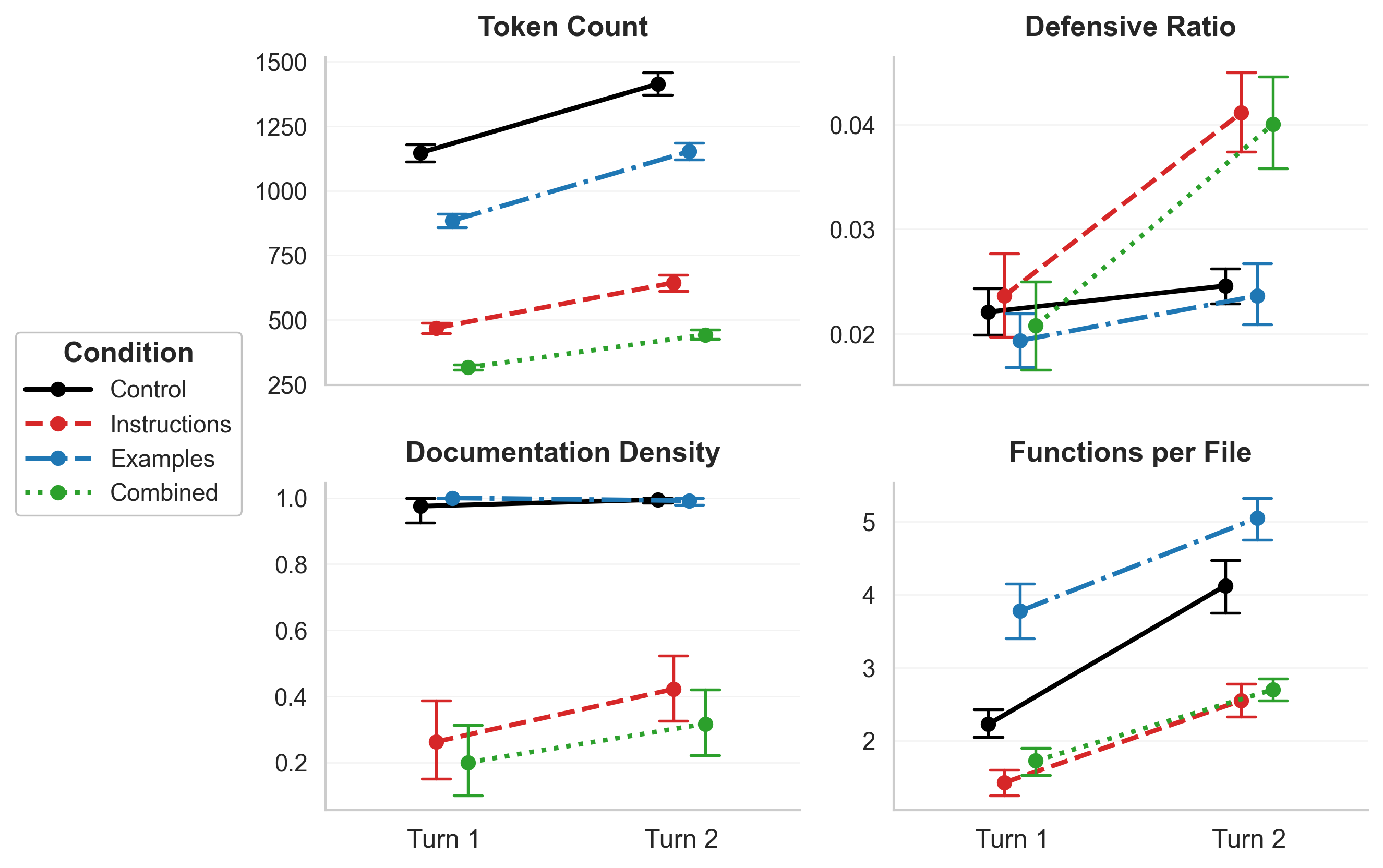}
\caption{Turn 1 to Turn 2 Trajectories (mean $\pm$ 95\% CIs).}
\label{fig:slopes}
\end{figure}

The revision prompt produced consistent increases in code length and defensive patterns regardless of initial system prompt, suggesting that generic revision instructions override condition-specific stylistic constraints.

\section{Discussion}\label{sec:discussion}

\subsection{Summary of Findings}\label{sec:disc-summary}
We tested whether instruction-based, example-based, and combined prompts differ in their effectiveness at controlling code style initially and in their ability to maintain compression discipline during feature expansion. Three research questions guided the analysis.

\textit{RQ1 (effectiveness):} How do instruction-based, example-based, and combined prompts differ in their initial stylistic control of verbosity relative to a control condition?
All three interventions reduced verbosity relative to Control, but magnitudes differed substantially. Combined produced strongest compression (d = -10.97, 70\% reduction), Instructions showed very large compression (d = -7.84, 56\% reduction), and Examples showed modest compression (d = -2.63, 20\% reduction). Functional accuracy remained near-ceiling across conditions. Explicit directives reshaped outputs most reliably; demonstrations alone permitted the model's default verbosity to partially persist. Combined succeeds by furnishing both concrete templates and explicit constraints, complementing broader evaluations of LLMs in software engineering that focus on correctness rather than format \citep{sarkar2025ai,zheng2025towards}.

\textit{RQ2 (mechanism):} Does combining instructions and examples produce additive, synergistic, or redundant effects on code style?
Effects are additive. Combined outperformed Instructions alone and Examples alone on initial verbosity. Secondary metrics reinforced this: Combined and Instructions elevated defensive ratio and reduced documentation density relative to Control, whereas Examples preserved documentation and showed neutral defensive patterns. Adding concrete demonstrations to explicit directives enhanced initial application without creating redundancy. Explicit directives exert stronger foundational control than demonstrations alone, and pairing them amplifies that control.

\textit{RQ3 (expansion discipline):} Which prompts maintain compression discipline when expanding functionality?
Directive-based prompts maintain compression discipline during expansion; example-based prompts do not. All conditions expanded when asked to enhance code for readability, maintainability, and error handling. However, expansion magnitude differed systematically. Combined added only 126 tokens during enhancement (140 fewer than Control, d = -1.51), Instructions added 175 tokens (92 fewer than Control, d = -0.85). Examples, despite starting 524 tokens more compact than Control, expanded at the same rate as Control ($\Delta_{\text{diff}} = +1.33$, d = 0.01). Initial compression and expansion discipline are governed by distinct mechanisms. Examples set a compact starting point but do not constrain growth rate; instructions establish rules that continue operating during expansion. Combined inherits directive discipline while retaining schema-anchoring benefits of examples.

Secondary metrics show selective rather than blanket constraint. Defensive ratio increased across all conditions in response to ``error handling'' requests, but directive-based conditions added checks more efficiently. Documentation density changed minimally. Functions per file increased universally in response to ``maintainability'' requests, but directive-based conditions moderated this expansion. Directive-based prompts respond to enhancement requests by adding necessary functionality while maintaining compression discipline, rather than rigidly resisting all expansion.

Functional correctness remained at ceiling across all condition-turn combinations ($>$97\%), establishing that verbosity differences reflect stylistic variation independent of functional capability, complementing evidence that LLM-generated code can pass tests while embedding subtle defects \citep{tambon2025bugs}.

\textit{Theoretical implications}
These findings distinguish initial effectiveness from expansion discipline as separate properties of prompting mechanisms. Examples yield strong first-pass compression through surface-form imitation; prior work shows example choice strongly shapes initial code generation and that models rely on format regularities in demonstrations rather than task semantics \citep{min2022rethinking,xu2024does}. Instruction-tuned models exhibit enhanced adherence to explicit directives, and those constraints persist when new functionality is requested \citep{ouyang2022training}. Combined's superior expansion discipline indicates demonstrations scaffold the application of directives, while demonstrations alone do not supply discipline during enhancement.

For single-shot generation, examples alone may suffice for modest compression. For iterative development where code must expand while remaining economical, instructions are essential. Optimal configuration pairs concrete examples that fix surface schemas with explicit rules that govern expansion behavior. This targets both immediate formatting and compression discipline during growth, relevant for multi-step, LLM-assisted development where agents support iterative refinement \citep{wang2025agents,laban2025llms}. Results complement work evaluating LLMs on design-pattern recognition \citep{pandey2025design} by showing low-level stylistic control can also be tuned systematically.

The divergence between Examples and directive-based conditions during expansion challenges a strong reading of few-shot prompting as establishing durable format constraints. If token-level schemas from exemplars governed subsequent behavior, carryover would appear during enhancement. Instead, influence appears confined to initial generation, consistent with evidence that demonstrations anchor surface patterns rather than persistent behavioral rules \citep{min2022rethinking}. Instructions persist because they function as testable constraints applicable across turns rather than surface templates applying only to immediate generation context \citep{ouyang2022training}. For multi-turn work, broader evaluations document reliability losses and constraint drift, underscoring why persistent rules matter \citep{laban2025llms,li2025beyond,hao2024empirical}.

\subsection{Turn 1: Initial Generation Effects}\label{sec:disc-turn1}
\textit{Instructions (d = -7.84).}
Directives alone produced strong stylistic control through specificity. Rather than abstract values, the prompt issued testable constraints about when to omit checks and docstrings and how to target minimality. This likely engaged instruction-following capacities emphasized during alignment \citep{ouyang2022training}. Results align with reports that models tend toward verbosity under weak constraint and that structured, task-specific instructions raise reliability while reducing tokens \citep{saito2023verbosity,cruz2025prompt,alhanahnah2025empirical,zheng2025towards}.

\textit{Examples (d = -2.63).}
Demonstrations alone delivered meaningful brevity gains but less than either directive-based condition. Behavior aligns with accounts of few-shot prompting as surface-form imitation: models borrow local schemas from exemplars and rely on format regularities in demonstrations rather than deeper task semantics \citep{min2022rethinking,xu2024does}. Examples achieved initial compression but matched Control on expansion rates, indicating templates set starting format but do not establish constraints governing subsequent growth.

\textit{Combined (d = -10.97).}
Combined paired explicit style directives with minimalist exemplars. Instructions fixed concrete targets for verbosity, documentation, and defensive structure; similar operational templates have increased pass rates while using fewer tokens \citep{cruz2025prompt}. Examples then instantiated these constraints, providing consistent structure the model could imitate, consistent with evidence that demonstrations steer surface patterns and that example choice substantially shapes code-generation quality \citep{min2022rethinking,xu2024does}. This pairing delivered the same functional behavior in roughly one third the tokens while keeping accuracy near ceiling, aligning with work showing developers iteratively refine prompts and examples to integrate assistants into collaborative development \citep{hao2024empirical}.

\textit{Secondary metrics.}
Combined and Instructions show higher defensive ratio despite large token reductions. This reflects a denominator effect: essential error handling remains while scaffolding shrinks, so checks form a larger share of a smaller program. Prompt design shifts multiple code-quality dimensions: Copilot prompts alter correctness, complexity, size, and similarity to human code \citep{fagadau2024analyzing}, and style-guiding prompts induce trade-offs between readability, conciseness, robustness, and functional accuracy \citep{wang2025beyond}.

Directive-based conditions sharply reduced documentation density, consistent with ``no docstrings unless necessary,'' whereas Examples preserved Control-level narration, in line with RLHF-tuned models exhibiting bias toward verbose responses when unconstrained \citep{ouyang2022training,saito2023verbosity}. Modularity moved in opposite directions: Examples increased functions per file, consistent with models imitating exemplars' compositional structure \citep{min2022rethinking}, while Instructions and Combined reduced decomposition, reflecting a pragmatic interpretation of ``minimum code'' favoring inline implementations over architectural elegance. These shifts mirror findings that prompt features alter structure alongside size and correctness, and that coding-style evaluations reveal trade-offs among readability, conciseness, and robustness \citep{fagadau2024analyzing,wang2025beyond,della2025prompt}.

\textit{Ordering and mechanism summary.}
Combined $>$ Instructions $>$ Examples. Demonstrations anchor local schema through surface-form imitation, whereas directives establish constraints that generalize beyond immediate generation context \citep{min2022rethinking,ouyang2022training}. Pairing a compact exemplar fixing surface schema with concrete rules establishes compression discipline for both initial generation and later enhancement, targeting first-pass formatting and expansion discipline during growth—the regime that matters for maintenance workflows where code evolves from MVP to production. Findings align with evidence that style-guideline prompting induces trade-offs between readability, conciseness, robustness, and accuracy \citep{fagadau2024analyzing,wang2025beyond,della2025prompt}.

\subsection{Turn 2: Expansion Discipline Under Enhancement Pressure}\label{sec:disc-turn2}
All conditions expanded when asked to enhance code for readability, maintainability, and error handling (all $d > 1.8$, all $p < .001$). Control increased +266 tokens ($d = 2.23$), Examples +268 ($d = 2.64$), Instructions +175 ($d = 1.84$), and Combined +126 ($d = 2.30$). This reflects realistic development pressure: enhancing code from minimal implementations to production-ready systems inherently adds tokens. The question is not whether expansion occurs, but which strategies maintain compression discipline during necessary growth.

Expansion magnitude differed systematically, $F(3, 156) = 21.51$, $p < .001$. Combined added 140 fewer tokens than Control ($d = -1.51$, $p < .001$), Instructions added 92 fewer ($d = -0.85$, $p < .001$), whereas Examples expanded at the same rate as Control ($\Delta_{\text{diff}} = +1.33$, $p = .957$, $d = 0.01$). The separation is stark: demonstrations delivered absolute brevity at Turn 1 (-524 tokens vs. Control) but provided no discipline during expansion; directive-based prompts maintained compression discipline throughout enhancement. Initial brevity is not a proxy for expansion discipline; constraints must remain operative when new functionality is requested.

Examples' failure reveals limits of surface-form imitation as a control mechanism. Despite starting 524 tokens more compact than Control, Examples expanded at identical rates, suggesting initial schema anchoring does not translate into constraints on growth. Two mechanisms plausibly explain this. First, enhancement requests introduce competing objectives, and in multi-turn settings such shifts associate with compounding assumptions and control drift \citep{laban2025llms,li2025beyond}. Second, token-level schemas from exemplars may anchor immediate generation without establishing decision rules applying to subsequent enhancement, consistent with evidence that in-context demonstrations shape local surface patterns rather than persistent behavioral rules \citep{min2022rethinking}.

By contrast, directive-based prompts maintained compression discipline. Explicit constraints remained operative even when enhancement objectives were introduced, consistent with instruction-following capacities applying operational rules across contexts rather than imitating surface patterns \citep{ouyang2022training}. Combined's stronger compression discipline relative to Control and advantage over Instructions indicates examples can reinforce how constraints are applied during expansion without supplying discipline independently. Demonstrations guide implementation via surface-form imitation, and few-shot code studies show example choice strongly shapes structure and content of generated programs \citep{min2022rethinking,xu2024does}. This division of labor explains why Combined outperforms both components: Instructions establish ``minimize bloat'' constraint, Examples provide templates for efficient implementation.

\textit{Secondary metrics reveal selective rather than blanket resistance.} Defensive ratio increased across all conditions in response to ``error handling'' requests, but more in Combined ($+0.0193$, $d = 0.83$, $p < .001$) and Instructions ($+0.0175$, $d = 0.82$, $p < .001$) than Examples ($+0.0043$, $d = 0.50$, $p = .003$) or Control ($+0.0025$, $d = 0.41$, $p = .014$), with between-condition differences significant, $F(3, 156) = 11.01$, $p < .001$. This reflects targeted enhancement rather than undisciplined elaboration: directive-based conditions responded by adding necessary error handling while maintaining overall compression discipline.

Documentation density changed minimally and inconsistently (Instructions $+0.159$, $d = 0.37$, $p = .024$; Combined $+0.117$, $d = 0.28$, $p = .087$; Control $+0.020$, $d = 0.12$, $p = .440$; Examples $-0.008$, $d = -0.23$, $p = .160$; between-condition $F(3, 156) = 2.57$, $p = .056$). ``Readability'' requests did not trigger substantial documentation additions, suggesting Turn 1 patterns persisted. Modularity increased in all conditions—Control +1.90 functions ($d = 1.42$, $p < .001$), Examples +1.27 ($d = 1.29$, $p < .001$), Instructions +1.12 ($d = 1.20$, $p < .001$), Combined +0.97 ($d = 1.48$, $p < .001$)—with directive-based conditions moderating expansion relative to Control (Examples $-0.62$, $d = -0.53$, $p = .020$; Instructions $-0.78$, $d = -0.67$, $p = .004$; Combined $-0.92$, $d = -0.88$, $p < .001$). ``Maintainability'' requests led all conditions to increase decomposition, but directive-based prompts moderated this tendency, balancing modularity gains against compression discipline. These metric-specific patterns suggest style dimensions respond independently and can trade off, consistent with evidence that style guidance and prompt features improve some qualities while increasing length or complexity \citep{wang2025beyond,fagadau2024analyzing}.

Differential expansion behavior establishes that initial effectiveness and expansion discipline are distinct properties requiring different mechanisms. Examples achieve initial compression through schema imitation but do not establish rules constraining subsequent growth. Instructions establish operational constraints—``write the minimum code that works''—functioning as decision rules applicable across enhancement contexts. Combined leverages both: examples scaffold efficient implementations of new features, while instructions maintain compression discipline preventing feature additions from bloating the codebase. For practitioners developing code evolving through multiple enhancement cycles, directive-based prompts are essential, and pairing them with compact examples optimizes both initial format and growth discipline. These findings align with analyses of LLM-based agents in software engineering highlighting importance of role design and structured multi-step tool use \citep{wang2025agents}, and with empirical studies showing multi-turn developer--ChatGPT conversations are common in collaborative coding \citep{hao2024empirical}.

\subsection{Limitations and Boundary Conditions}\label{sec:disc-limitations}
Our conclusions concern stylistic control in a two-turn code-generation protocol with near-ceiling functional accuracy; they do not generalize to task success per se. We compare four prompt conditions but do not estimate factorial interactions, so Combined's advantage is descriptive rather than a formal interaction effect. Results are conditional on one model family and version, the \texttt{cl100k\_base} tokenizer, and fixed sampling and seed settings; different models, inference regimes, or tokenizers may change effect magnitudes and condition rankings. The prompts instantiate a single directive set and exemplar set, and expansion discipline is assessed for one enhancement request; alternative wordings, exemplars, or objectives could yield different dynamics, consistent with evidence that post-training procedures can increase talkativeness and impose a ``verbose tax'' under weaker constraints \citep{sun2023principle}. Functional evaluation is deliberately minimal: a single binary test on a fixed dummy dataset isolates stylistic differences but cannot capture defect modes, repair opportunities, or partial correctness. Empirical work on LLM-generated defects shows that seemingly plausible code can still harbor subtle errors \citep{tambon2025bugs}, and recent work on test selection for LLMs demonstrates that coverage and test choice strongly shape measured reliability \citep{quan2025evaluation}. Our results therefore speak to style conditional on a narrow notion of correctness and would be usefully complemented by evaluations that vary test coverage, incorporate mutation-based testing, or couple style prompts with explicit fault models. Similarly, we do not analyze higher-level design structures; findings from design-pattern recognition with LLMs indicate that structural semantics can be captured and manipulated, which lies outside our token- and ratio-based metrics \citep{pandey2025design}.

Our metrics emphasize token counts and simple structural ratios rather than human judgments of readability or maintainability. We foreground effect sizes with unadjusted two-sided p-values and treat omnibus tests as descriptive summaries of pattern strength. Within these scope conditions, the central result holds: concrete directives yield stronger initial stylistic control and better expansion discipline than demonstrations alone. The applicability of this result to broader practice depends on how LLM coding tools are embedded into pipelines that include repositories, testing infrastructure, and human review. Field deployments and agent-style workflows introduce additional constraints, tool integrations, and organizational practices that may interact with prompt-level style control, as suggested by studies of developer--ChatGPT conversations, agentic productivity tools, and AI-mediated software-development processes \citep{hao2024empirical,sarkar2025ai,sauvola2024future}.

\subsection{Future Directions}\label{sec:disc-future}
Our findings show that directive-based prompts maintain compression discipline during enhancement while example-based prompts do not, but the underlying mechanism remains unresolved. We attribute directive persistence to instruction-following capacities emphasized during alignment, yet this is descriptive rather than explanatory \citep{ouyang2022training}. Why does ``write the minimum code that works'' remain operative across enhancement cycles while concrete examples anchor only initial generation? Plausible mechanisms include differential attention over long inputs that weakens retention of earlier guidance, stronger adherence to explicit rules due to instruction tuning, and surface-pattern imitation from exemplars that lacks persistent decision rules \citep{min2022rethinking,liu2024lost,li2025beyond}. Experiments that manipulate directive specificity and exemplar variety while probing internal model states would clarify which mechanisms drive expansion discipline.

Whether compression discipline produces code that human developers actually prefer for maintenance also remains an open question. Our metrics indicate that directive-based prompts reduce tokens while increasing defensive ratio and moderating modularity, but it is unclear whether developers experience this compressed code as easier to understand, modify, and extend. Aggressive minimization could sacrifice clarity that matters for collaboration \citep{fagadau2024analyzing,wang2025beyond}. Pairing style metrics with expert assessment, studies of collaborative ChatGPT usage in real repositories \citep{hao2024empirical}, and longer-run evaluations of maintainability would determine whether compression discipline optimizes the right outcomes or simply minimizes token count.

Finally, we evaluated compression discipline over a single enhancement cycle, whereas real codebases evolve through repeated changes with shifting priorities such as hardening, feature addition, and refactoring \citep{sergeyuk2025using}. Surveys and field studies indicate that AI assistants and agents are used unevenly across lifecycle stages and are increasingly embedded in multi-agent configurations \citep{wang2025agents,sauvola2024future,sarkar2025ai}. Sequential enhancement protocols would reveal whether directive-based constraints persist or decay over multiple cycles, consistent with open challenges in sustaining behavior across multi-turn interactions \citep{li2025beyond}. Combining style-oriented prompting with architectural supports such as RAG-enhanced coding agents, declarative specifications, or dedicated repair agents could enable workflows where one agent generates code and another enforces compression discipline or performs targeted repair \citep{he2026enhancing,alhanahnah2025empirical}.

\section{Conclusion}\label{sec:conclusion}
This study asked whether abstract directives, concrete exemplars, or their combination provide reliable style control in LLM code generation and whether that control endures during enhancement. Across 320 programs, all prompts reduced initial verbosity while keeping functional accuracy near ceiling; the Combined condition yielded the strongest compression (about $-70\%$ tokens vs. Control), with Instructions close behind and Examples providing a modest reduction. Directive-based prompts also maintained greater compression discipline during revision, adding substantially fewer tokens than Control, whereas Examples expanded at control-like rates. These results separate two properties of prompt design: initial effectiveness (how compact first-pass code is) and expansion discipline (how tightly style is held when functionality grows). Conceptually, the findings reinforce that style control is distinct from correctness and that durable control depends on explicit operational rules rather than template imitation; methodologically, the paired two-turn protocol and compact style metrics offer a practical evaluation pattern for future work. For iterative workflows, the actionable guidance is straightforward: pair concise examples with concrete directives to anchor format and preserve economy through revision. Taken together, disciplined prompting supplies a tractable lever for predictable code style without sacrificing task success.

\section*{Statements and Declarations}

\bmhead{Funding}
No funds, grants, or other support was received for conducting this study.

\bmhead{Competing interests}
The author declares no competing interests.

\bmhead{Data availability}
All data, code, and analysis scripts are available at \url{https://github.com/jeremiahbohr/show-and-tell}.

\bmhead{Author contributions}
J.B. conceived and designed the study, collected and analyzed the data, and wrote the manuscript.

\appendix
\section{Appendix}\label{sec:appendix}

\subsection*{Examples System Prompt}

\begin{verbatim}
SYSTEM_PROMPT = """
Good code example:
```python
def parse_csv(path):
    return pd.read_csv(path)

def calc_stats(df):
    return {
        'total': len(df),
        'unique_users': df['user_id'].nunique(),
        'avg_by_type': df.groupby('event_type')['value'].mean().to_dict()
    }
```
"""
\end{verbatim}

\subsection*{Instructions System Prompt}

\begin{verbatim}
SYSTEM_PROMPT = """
I value minimal, functional code. No defensive coding unless explicitly required. 
No docstrings unless function purpose is non-obvious from the name and signature.

Write the minimum code that works.
"""
\end{verbatim}

\subsection*{Combined System Prompt}

\begin{verbatim}
SYSTEM_PROMPT = """
I value minimal, functional code. No defensive coding unless explicitly required. 
No docstrings unless function purpose is non-obvious from the name and signature.

Good code example:
```python
def parse_csv(path):
    return pd.read_csv(path)

def calc_stats(df):
    return {
        'total': len(df),
        'unique_users': df['user_id'].nunique(),
        'avg_by_type': df.groupby('event_type')['value'].mean().to_dict()
    }
```

Write the minimum code that works.
"""
\end{verbatim}

\bibliography{references}

\end{document}